\documentclass[prd,superscriptaddress,amsfonts,amssymb,amsmath,showpacs,onecolumn]{revtex4-2}

\usepackage{bm}
\usepackage{amsfonts}
\usepackage{latexsym}
\usepackage[latin1]{inputenc}
\usepackage{graphicx}
\usepackage{amsmath}
\usepackage{palatino}
\usepackage{mathpazo}
\usepackage{textcomp}
\usepackage{float}
\usepackage{booktabs}
\usepackage{dcolumn}
\usepackage{ragged2e}
\usepackage{hyperref}
\hypersetup{colorlinks,citecolor=blue}
\hypersetup{colorlinks=true,linkcolor=red,filecolor=magenta,    urlcolor=blue}
\usepackage{amsmath}
\usepackage{graphicx}% Include figure files
\usepackage{makecell}
\usepackage{hyperref}% add hypertext capabilities
\usepackage[mathlines]{lineno}% Enable numbering of text and display math
\usepackage{amssymb}
\usepackage{enumitem}
\usepackage{mathtools}
\usepackage{mathrsfs}
\usepackage{setspace}
\usepackage{color}
\usepackage{graphicx}
\usepackage[dvipsnames]{xcolor}
\usepackage{mathalfa}
\usepackage{calligra}
\usepackage[misc]{ifsym}
\DeclareMathAlphabet{\mathpzc}{OT1}{pzc}{m}{it}
\DeclareMathAlphabet{\mathcalligra}{T1}{calligra}{m}{n}
\hypersetup{colorlinks=true,citecolor=blue,linkcolor=red,urlcolor=magenta}
\usepackage{xcolor}
\usepackage{orcidlink}
\usepackage[caption=false]{subfig}
\usepackage{commath}
\def\jnl@style{}
\def\aaref@jnl#1{{\jnl@style#1}}

\def\aaref@jnl#1{{\jnl@style#1}}

\def\aj{\aaref@jnl{AJ}}                   % Astronomical Journal
\def\apj{\aaref@jnl{ApJ}}                 % Astrophysical Journal
\def\apjl{\aaref@jnl{ApJ}}                % Astrophysical Journal, Letters
\def\apjs{\aaref@jnl{ApJS}}               % Astrophysical Journal, Supplement
\def\apss{\aaref@jnl{Ap\&SS}}             % Astrophysics and Space Science
\def\aap{\aaref@jnl{A\&A}}                % Astronomy and Astrophysics
\def\aapr{\aaref@jnl{A\&A~Rev.}}          % Astronomy and Astrophysics Reviews
\def\aaps{\aaref@jnl{A\&AS}}              % Astronomy and Astrophysics, Supplement
\def\mnras{\aaref@jnl{Mon.~Not.~Roy.~Astron.~Soc.}}             % Monthly Notices of the RAS
\def\prd{\aaref@jnl{Phys.~Rev.~D}}        % Physical Review D
\def\plb{\aaref@jnl{Phys.~Lett.~B}}        % Physics Letters B
\def\prc{\aaref@jnl{Phys.~Rev.~C}}  % Physical Review C
\def\prl{\aaref@jnl{Phys.~Rev.~Lett.}}    % Physical Review Letters
\def\qjras{\aaref@jnl{QJRAS}}             % Quarterly Journal of the RAS
\def\skytel{\aaref@jnl{S\&T}}             % Sky and Telescope
\def\ssr{\aaref@jnl{Space~Sci.~Rev.}}     % Space Science Reviews
\def\zap{\aaref@jnl{ZAp}}                 % Zeitschrift fuer Astrophysik
\def\nat{\aaref@jnl{Nature}}              % Nature
\def\aplett{\aaref@jnl{Astrophys.~Lett.}} % Astrophysics Letters
\def\apspr{\aaref@jnl{Astrophys.~Space~Phys.~Res.}} % Astrophysics Space Physics Research
\def\physrep{\aaref@jnl{Phys.~Rep.}}      % Physics Reports
\def\physscr{\aaref@jnl{Phys.~Scr}}       % Physica Scripta
\def\commat{\aaref@jnl{Comm.~Math.~Phys.}}              % Communications in Mathematical Physics
\def\science{\aaref@jnl{Science}}               % Science
\def\cqg{\aaref@jnl{Classical Quant.~Grav.}}            % Classical and Quantum Gravity
\def\jpcs{\aaref@jnl{JPCS}}                                     % Journal of Physics Conference Series
\def\ijmpd{\aaref@jnl{Int.~J.~Mod.~Phys.~D}}                    % International Journal of Modern Physics D
\def\grg{\aaref@jnl{Gen.~Relat.~Gravit.}}               % General Relativity and Gravitation
\def\rpp{\aaref@jnl{Rep.~Prog.~Phys.}}          % Reports on Progress in Physics
\def\npa{\aaref@jnl{Nucl.~Phys.~A}}        % Nuclear Physics A
\def\lrr{\aaref@jnl{Living Rev.~Rel.}}                   % Living reviews in relativity
\def\jcap{\aaref@jnl{J.~Cosmology Astropart.~Phys.}}    % Journal of cosmology and astroparticle physics
\def\rmp{\aaref@jnl{Rev.~Mod.~Phys.}}   %Reviews of modern physics
\def\epjc{\aaref@jnl{Eur.~Phys.~J.~C}}

\allowdisplaybreaks[1]
\renewcommand{\arraystretch}{1.1}
\begin{document}

\preprint{APS/123-QED}

\title{Conformally symmetric wormhole solutions supported by non-commutative geometry in $\mathpzc{f}( \mathcal{Q}, \mathcal{T})$ gravity }% Force line breaks with \\

\author{Chaitra Chooda Chalavadi\orcidlink{0000-0003-1976-2307}}
\email{chaitra98sdp@gmail.com} 
\affiliation{Department of P.G. Studies and Research in Mathematics,
 \\
 Kuvempu University, Shankaraghatta, Shivamogga 577451, Karnataka, INDIA
}%

\author{V. Venkatesha\orcidlink{0000-0002-2799-2535}}%
 \email{vensmath@gmail.com}
\affiliation{Department of P.G. Studies and Research in Mathematics,
 \\
 Kuvempu University, Shankaraghatta, Shivamogga 577451, Karnataka, INDIA
}%

\author{N. S. Kavya\orcidlink{0000-0001-8561-130X}}
\email{kavya.samak.10@gmail.com}
\affiliation{Department of P.G. Studies and Research in Mathematics,
 \\
 Kuvempu University, Shankaraghatta, Shivamogga 577451, Karnataka, INDIA
}% 

\author{S. V. Divya Rashmi}
\email{rashmi.divya@gmail.com}
\affiliation{Department of Mathematics, Vidyavardhaka College of Engineering,\\
Mysuru - 570002, INDIA}



\date{\today}% It is always \today, today,

\begin{abstract}
This manuscript investigates wormhole solutions within the framework of extended symmetric teleparallel gravity, incorporating non-commutative geometry, and conformal symmetries. To achieve this, we examine the linear wormhole model with anisotropic fluid under Gaussian and Lorentzian distributions. The primary objective is to derive wormhole solutions while considering the influence of the shape function on model parameters under Gaussian and Lorentzian distributions. The resulting shape function satisfies all the necessary conditions for a traversable wormhole. Furthermore, we analyze the characteristics of the energy conditions and provide a detailed graphical discussion of the matter contents via energy conditions. Additionally, we explore the effect of anisotropy under Gaussian and Lorentzian distributions. Finally, we present our conclusions based on the obtained results.
\begin{description}
\item[Keywords]
Traversable wormhole, f(Q,T) gravity, energy conditions, non-commutative geometry,\\ conformal motion
  
\end{description}
\end{abstract}

\maketitle

%\tableofcontents
\section{Introduction}\label{section I}
Wormholes are dual-mouthed hypothetical structures connecting distinct sectors in the same universe or different universes. Initially, Flamm \cite{flamm} introduced the notion of a wormhole by constructing the isometric embedding of the Schwarzschild solution. Einstein and Rosen \cite{einstein} employed Flamm's concept to create a bridge, commonly known as the Einstein-Rosen bridge. Later, Thorne and his student Morris \cite{thorne} conducted pioneering research on the concept of traversable wormholes. They meticulously examined static and spherically symmetric wormholes, revealing that the exotic matter inside them possess negative energy, thus violating the null energy condition. Furthermore, in order to establish a physically feasible model, it is necessary to repudiate the existence of the hypothetical matter. Although within the framework of general relativity \cite{gonza,armend}, it was not possible to definitively rule out the presence of such a substance, an alternative approach was supported to reduce or eliminate the reliance on exotic matter \cite{visser,visse,kuhfitting}. Numerous studies have been conducted to explore wormhole solutions within the background of modified theories \cite{franc,azizi,bohmer,sharif,singh,jimenez,yixin,rahaman,zubair,maldacena,ovgun,mustafa,elizalde,anchor,lobo,capozzi,capozz,chaitra,ad1,ad2,kavya3,ad3,ad4,kavya4,ad5,ad6,gm1,gm2}.

 In the context of string theory, non-commutative geometry is one of the most intriguing concepts. The idea of non-commutativity arises from the notion that coordinates on a D-brane can be treated as non-commutative operators. This property holds great significance in mathematically explored fundamental concepts of quantum gravity \cite{doplicher,witten,seiberg}. Non-commutative geometry aims to unify spacetime gravitational forces with weak and strong forces on a single platform. Within this framework, it becomes possible to replace point-like structures with smeared objects, leading to the discretization of spacetime. This discretization arises from the commutator $[x^a, x^b]= i \theta^{ab}$ where $\theta^{ab}$ is an antisymmetric second-order matrix \cite{kase,smailagic,nicolini}. To simulate this smearing effect, the Gaussian distribution and  Lorentzian distribution with a minimum length of $\sqrt{\theta}$ are incorporated instead of the Dirac delta function. This non-commutative geometry is an intrinsic property of space-time and independent of the behavior of curvature.
 
 Non-commutative geometry plays a crucial role in examining the properties of space-time geometry under different conditions. Jamil et al., \cite{jamil} explored some new exact solutions of static wormholes under non-commutative geometry.  They utilized the power-law approach to analyze these solutions and discuss their properties. Rahaman et al. \cite{arxiv, baner, farook} conducted an extensive investigation into various studies in non-commutative geometry. They studied fluids in different dimensions influenced by non-commutative geometry, which exhibited conformal symmetry. Additionally, they derived specific solutions of a wormhole within the context of $\mathpzc{f}(\mathcal{R})$ gravity. 
 In the realm of non-commutative geometry, Zubair et al. \cite{waheed} examined wormhole solutions that permit conformal motion within the context of $\mathpzc{f}(\mathcal{R}, \mathcal{T})$ theory. The study employed conformal killing vectors to analyze the properties and characteristics of these wormhole solutions. Kuhfitting \cite{indian} investigated the stable wormhole solutions utilizing conformal killing vectors within the framework of a non-commutative geometry that incorporates a minimal length. The study focused on exploring the properties and characteristics of these stable wormholes within this specific theoretical framework. In \cite{shamir}, the authors studied the non-commutative wormhole solution in $\mathpzc{f}(\mathcal{R})$ gravity. Moreover, the concept of non-commutative geometry has been gaining attention from researchers, and numerous intriguing aspects of this theory have been extensively explored and deliberated upon in the literature \cite{aschieri,schne,sushkov,high,garatt,class,energy,cosmo,math,rahaman2,rahaman3,gm3,zinnat,kavya, kavya2}. Inspired by the aforementioned attempts in modified gravity and non-commutative geometry, we now delve into the study of wormhole solutions in $\mathpzc{f}(\mathcal{Q}, \mathcal{T})$ gravity. We consider Gaussian and Lorentzian non-commutative geometries with conformal killing vectors to explore their implications.
   
   The manuscript is structured following the subsequent pattern: In section \ref{sectionII}, we discuss the traversability condition for a wormhole. We shall construct the mathematical formalism of $\mathpzc{f}(\mathcal{Q}, \mathcal{T})$ gravity in \ref{sectionIII}. In the same section, we briefly explain the energy condition and the basic formalism of conformal killing vectors. In section \ref{section IV}, we conduct a detailed analysis of the wormhole model under Gaussian and Lorentzian distributions. Within this section, we derive the shape function and explore the impact of model parameters on these functions, as well as the energy conditions. In section \ref{section V}, we investigate the effect of anisotropy on both distributions. Finally, in section \ref{section VI}, we finalized the conclusive remarks and summarized the key findings of the study.
	
%%%%%%%%%%%%%%%%%%%%%%%%%%%%%%%%%%%%%%%%%%%%%%%%%%%%%%%%%%%%%%%%%%%%%%%%%%%
\section{TRAVERSABILITY CONDITIONS FOR WORMHOLE}\label{sectionII}
	The Morris-Thorne metric for the traversable wormhole is described as
	
		\begin{equation}\label{whmetric}
			ds^2=e^{2\Phi(r)}dt^2-\dfrac{dr^2}{1-\dfrac{\Psi(r)}{r}  } - r^2\left(d\theta^2+\text{sin}^2\theta \,d\phi^2\right).
		\end{equation}  
	In this scenario, we have two functions, namely $\Phi(r)$ and $\Psi(r)$  which are referred to as the redshift and shape functions respectively. Both of these functions depend on the radial coordinate $r$.
		\begin{enumerate}[label=$\arabic*.$,leftmargin=*]
			\setlength{\itemsep}{4pt}
			\setlength{\parskip}{4pt}
			\setlength{\parsep}{4pt}
               \item Redshift function:  The redshift function $\Phi(r)$ needs to have a finite value across the entire space-time. Additionally, the redshift function must adhere to the constraint of having no event horizon, which allows for a two-way journey through the wormhole.
              
               \item Shape function: The shape function $\Psi(r)$ characterizes the geometry of the traversable wormhole. Therefore, $\Psi(r)$  must satisfy the following conditions:
	 \begin{enumerate}[label=$\bullet$,leftmargin=*]
			\setlength{\itemsep}{4pt}
			\setlength{\parskip}{4pt}
			\setlength{\parsep}{4pt}

 \item \textit{Throat condition}: The value of the function $\Psi(r)$ at the throat is $r_0$ and hence $1-\frac{\Psi(r)}{r}>0$ for $r>r_0.$ 
			\item \textit{Flaring-out condition:} The radial differential of the shape function, $\Psi'(r)$ at the throat should satisfy, $\Psi'(r_0)<1.$ 
			\item \textit{Asymptotic Flatness condition:} As $r\rightarrow \infty$, $\frac{\Psi(r)}{r}\rightarrow 0$.
		\end{enumerate}
  
		\item Proper radial distance function: This function should be finite everywhere in the domain. In magnitude, it decreases from the upper universe to the throat and then increases from the throat to the lower universe. The proper radial distance function is expressed as,
	
		\begin{equation}\label{proper}
			\mathit{\mathit{l}}(r)=\pm \int_{r_0}^r \dfrac{dr}{\sqrt{\dfrac{r-\Psi(r)}{r}}}.
		\end{equation}
	\end{enumerate}	
		
%%%%%%%%%%%%%%%%%%%%%%%%%%%%%%%%%%%%%%%%%%%%%%%%%%%%%%%%%%%%%%%%%
\section{ Mathematical formulations of $\mathpzc{f}(\mathcal{Q},\mathcal{T})$ GRAVITY }\label{sectionIII}	
	In this article, we are particularly interested in $\mathpzc{f}(\mathcal{Q},\mathcal{T})$ gravity, where the Lagrangian is an arbitrary function of non-metricity scalar and the trace of the energy-momentum tensor. Yixin et al., \cite{yixin} introduced the $\mathpzc{f}(\mathcal{Q},\mathcal{T})$ gravity, which is referred to as extended symmetric teleparallel gravity. This was developed within the metric-affine formalism framework. $\mathpzc{f}(\mathcal{Q},\mathcal{T})$ gravity theory has been employed to explain both matter-antimatter asymmetry and late-time acceleration. Furthermore, recent investigations suggest that $\mathpzc{f}(\mathcal{Q},\mathcal{T})$ gravity may provide a feasible explanation of various cosmological and astrophysical phenomena \cite{bhatta,arora,chaitra,chaitra2}. Nevertheless, no further studies on wormholes were conducted based on this theory, which is still in its early stages of development. These considerations motivate us to select the $\mathpzc{f}(\mathcal{Q},\mathcal{T})$ gravity to derive wormhole solutions.
 
  The Einstein Hilbert action for $\mathpzc{f}(\mathcal{Q},\mathcal{T})$ gravity is given by
		\begin{equation}\label{action}
			S=\int \frac{1}{16\pi}\, \mathpzc{f}(\mathcal{Q},\mathcal{T})	\sqrt{-g}\, d^4x +\int \mathscr{L}_m \sqrt{-g}\, d^4x,
		\end{equation}
		where $\mathpzc{f}(\mathcal{Q},\mathcal{T})	$ is an arbitrary  function that couples the non-metricity  $\mathcal{Q}$ and the trace $\mathcal{T}$ of the energy momentum tensor, $\mathscr{L}_m$ is the Lagrangian density corresponding to matter and $g$ denote the determinant of the metric $g_{\mu\nu}$. 
                               
	The Non-metricity tensor is defined as
	    \begin{equation}\label{nonmetricity}
	        \mathcal{Q}_{\lambda\mu\nu} = \nabla_\lambda g_{\mu\nu},
	    \end{equation}
    and its traces are
	    \begin{equation}
	        \mathcal{Q}_\alpha = \mathcal{Q}^{\;\;\;\mu}_{\alpha\;\;\mu}, \quad \Tilde{\mathcal{Q}}_\alpha = \mathcal{Q}{^\mu}_{\alpha\mu}.
	    \end{equation}
     
	   Further, we can define a super-potential associated with the non-metricity tensor as
	    	    \begin{equation}\label{sp}
	         {P^\alpha} _{\mu\nu} = \frac{1}{4}\left[-\mathcal{Q}^\alpha_{\;\;\mu\nu} + 2\mathcal{Q}^{\;\;\;\alpha}_{\left(\mu\;\;\nu\right)}+ \mathcal{Q}^\alpha g_{\mu\nu} -\Tilde{\mathcal{Q}}^\alpha g_{\mu\nu} -\delta^\alpha_{(\mu}\mathcal{Q}_{\nu)}\right],
	    \end{equation}
	    
	   The non-metricity scalar is represented as
	    \begin{equation}
	        \mathcal{Q} = -\mathcal{Q}_{\alpha \mu \nu} P^{\alpha \mu \nu}.
		 \end{equation}

Besides, the energy-momentum tensor for the fluid depiction of space-time can be expressed as
		\begin{equation}
		   \mathcal{T}_{\mu \nu} = \frac{-2}{\sqrt-g} \frac{\delta(\sqrt -g \mathscr{L}_m)}{\delta g^{\mu \nu}},
		\end{equation}
		and
		\begin{equation} \label{tm}
		    \Theta _{\mu \nu} = g^{\alpha \beta} \frac{\delta \mathcal{T}_{\alpha \beta}}{\delta g^{\mu \nu}}.
		\end{equation}

The variation of the action \eqref{action} with respect to the fundamental metric, gives the metric field equation 
	   \begin{equation}\label{fd}
		     \frac{-2}{\sqrt-g} \nabla_\alpha \left(\sqrt -g \mathpzc{f}_\mathcal{Q} P^\alpha _{\;\;\mu \nu}\right) - \frac{1}{2} g_{\mu\nu} \mathpzc{f} + \mathpzc{f}_\mathcal{T}\left( \mathcal{T}_{\mu \nu} + \Theta_{\mu \nu}\right)  - \mathpzc{f}_\mathcal{Q} \left(P_{\mu \alpha \beta} \mathcal{Q}^{\;\;\alpha \beta} _{\nu} -2\mathcal{Q}^{\alpha \beta} _{\;\;\mu} P_{\alpha \beta\nu}\right) = 8\pi \mathcal{T}_{\mu\nu}.
		\end{equation}
		where $\mathpzc{f}_\mathcal{Q} = \frac{d\mathpzc{f}}{d\mathcal{Q}}$ and $\mathpzc{f}_\mathcal{T} = \frac{d\mathpzc{f}}{d\mathcal{T}}$.
 
We presume that the matter distribution is an anisotropic stress-energy tensor, which can be written as
		\begin{equation}\label{energymomentumtensor}
			\mathcal{T}_{\mu\nu}=(\rho+p_t)\eta_\mu \eta_\nu-p_t\,g_{\mu\nu}+(p_r-p_t)\mathfrak{x}_{\mu} \mathfrak{x}_\nu,
		\end{equation}
		where $\rho, p_r, p_t$ are the energy density, radial and tangential pressures respectively. Here, $\eta_\mu$ refers to a four-velocity vector with a magnitude of one, while $\mathfrak{x}_\mu$ represents a space-like unit vector. Additionally, in this scenario, the tangential pressure will be orthogonal to the unit vector, and the radial pressure will be along the four-velocity vector.
  \vspace{25pt}\\
  The expression for the trace of the energy-momentum tensor is determined as $\mathcal{T}= \rho - p_r - 2p_t$ and equation \eqref{tm} can be read as
		\begin{equation}
		    \Theta _{\mu \nu} = -g_{\mu\nu}\,\dfrac{p_r+2p_t}{3}- 2\,\mathcal{T}_{\mu\nu}.
		\end{equation}

		Using the wormhole metric \eqref{whmetric}, the trace of the non-metricity scalar $\mathcal{Q}$ can be written as,
		\begin{equation}{\label{nms}}
		    \mathcal{Q} = -\frac{\Psi}{r^2} \left[ \frac{r\Psi'-\Psi}{r(r-\Psi)} + 2 \Phi'\right].
		\end{equation}

  Now, substituting the wormhole metric \eqref{whmetric} and anisotropic matter distribution \eqref{energymomentumtensor} into the motion equation \eqref{fd}, we found the following expressions:
  \begin{equation}
      \label{fe1}
      \begin{split}
          \rho= -\dfrac{1}{48\pi r^3 (r-\Psi) \left(\mathpzc{f}_{\mathcal{T}} + 8\pi\right)} \Bigg[&\mathpzc{f}_{\mathcal{Q}} \mathpzc{f}_{\mathcal{T}} \Biggl(-r\Psi' \left(2r(r-\Psi)\Phi' + \Psi + 2r\right)+ 3\Psi^2 \Biggl.\Bigg.\\
          & \Biggl.\Bigg.
          +4r(\Psi-r) \left(r(\Psi-r)\Phi'' + \Phi' \left(r (\Psi-r)\Phi' +3\Psi-2r\right)\right)\Biggl)\Bigg.\\
          &\Bigg. +24 \pi \mathpzc{f}_{\mathcal{Q}}\left(\Psi\left(2r(\Psi-r)\Phi'+\Psi\right) + r(\Psi-2r)\Psi'\right)\Bigg.\\
          &\Bigg.
          +r(\Psi-r)\left(\mathpzc{f}_{\mathcal{T}}\left(2 \mathpzc{f}_{\mathcal{Q}\mathcal{Q}} \mathcal{Q}' \left(2r(\Psi-r)\Phi' +\Psi\right)+3\mathpzc{f} r^2\right)+ 48 \pi \Psi \mathpzc{f}_{\mathcal{Q}\mathcal{Q}} \mathcal{Q}'+ 24\pi \mathpzc{f} r^2\right)\Bigg],
      \end{split}
  \end{equation}

 \begin{equation}
     \label{fe2}
     \begin{split}
         p_r=\dfrac{1}{48\pi r^3 (r-\Psi) \left(\mathpzc{f}_{\mathcal{T}} + 8\pi\right)} \Bigg[&\mathpzc{f}_{\mathcal{Q}} \mathpzc{f}_{\mathcal{T}} \Biggl(r\Psi' \left(2r(r-\Psi)\Phi' + \Psi + 2r\right)- 3\Psi^2 \Biggl. \Bigg.\\
         &\Biggl.\Bigg.
         - 4r(\Psi-r) \left(r(\Psi-r)\Phi'' + \Phi' \left(r (\Psi-r)\Phi' +3\Psi-2r\right)\right)\Biggl)\Bigg.\\
         &\Bigg.+24\pi \mathpzc{f}_{\mathcal{Q}}\left(\Psi r\Psi'- (3\Psi-2r) \left(2r(\Psi-r)\Phi' +\Psi\right)\right)\Bigg.\\
         &\Bigg.-r(\Psi-r)\left(\mathpzc{f}_{\mathcal{T}}\left(2 \mathpzc{f}_{\mathcal{Q}\mathcal{Q}} \mathcal{Q}' \left(2r(\Psi-r)\Phi' +\Psi\right)+3\mathpzc{f} r^2\right)+ 48\pi \Psi \mathpzc{f}_{\mathcal{Q}\mathcal{Q}} \mathcal{Q}'+24\pi \mathpzc{f} r^2\right)\Bigg],
     \end{split}
 \end{equation}
		    
\begin{equation}
    \label{fe3}
    \begin{split}
        p_t=-&\dfrac{1}{48\pi r^3 (r-\Psi) \left(\mathpzc{f}_{\mathcal{T}} + 8\pi\right)} \Bigg[\mathpzc{f}_{\mathcal{Q}} \mathpzc{f}_{\mathcal{T}} \Biggl(-r\Psi' \left(2r(r-\Psi)\Phi' + \Psi + 2r\right)+ 3\Psi^2\Biggl.\Bigg.\\
        &\Biggl.\Bigg.+4r(\Psi-r) \left(r(\Psi-r)\Phi'' + \Phi' \left(r (\Psi-r)\Phi' +3\Psi-2r\right)\right)\Biggl)\Bigg.\\
        &\Bigg.+24\pi r\mathpzc{f}_{\mathcal{Q}} \left(r \left(2(\Psi-r)^2\left(\Phi'' + (\Phi')^2\right) + (2r - 5\Psi)\Phi' + \Psi'\left(\Phi' (\Psi-r)-1\right)\right) +\Psi(3\Psi\Phi' +1)\right)\Bigg.\\
        &\Bigg. +r(\Psi-r)\left(\mathpzc{f}_{\mathcal{T}}\left(2\Psi \mathpzc{f}_{\mathcal{Q}\mathcal{Q}} \mathcal{Q}'+3\mathpzc{f} r^2\right) + 4r(\Psi-r)\mathpzc{f}_{\mathcal{Q}\mathcal{Q}} \left(\mathpzc{f}_{\mathcal{T}} +12\pi\right) \mathcal{Q}'\Phi' +24\pi\mathpzc{f} r^2\right)\Bigg].
    \end{split}
\end{equation}
 
%%%%%%%%%%%%%%%%%%%%%%%%%%%%%%%%%%%%%%%%%%%%%%%%%%%%%%%%%%%%%%%%%%%%%%%%%%%%%%%%%%%%%%
\subsection{Energy Condition}
\par Energy conditions provide interpretations for the physical phenomena associated with the motion of energy and matter, which are derived from the Raychaudhuri equation. To evaluate the geodesic behavior, we shall consider the criterion for different energy conditions. For the anisotropic matter distribution with $\rho, p_r$ and $p_t$ being energy density, radial pressure and tangential pressure, we have the following:
		\begin{enumerate}[label=$\circ$,leftmargin=*]
			\setlength{\itemsep}{4pt}
			\setlength{\parskip}{4pt}
			\setlength{\parsep}{4pt}
			\item \textit{Null Energy Conditions}: $\rho+p_t\ge0$ and $\rho+p_r\ge0$.
			\item \textit{Weak Energy Conditions}: $\rho\ge0\implies$  $\rho+p_t\ge0$ and $\rho+p_r\ge0$.
			\item \textit{Strong Energy Conditions}: $\rho+p_j\ge0\implies$  $\rho+\sum_j p_j\ge0   \ \forall\ j$.
			\item \textit{Dominant Energy Conditions}: $\rho\ge0\implies$ $\rho-|p_r|\ge0$ and $\rho-|p_t|\ge0$.
		\end{enumerate}
%%%%%%%%%%%%%%%%%%%%%%%%%%%%%%%%     %%%%%%%%%%%%%%%%%%%%%%%%%%%%%%%%%     
\subsection{Conformal Killing Vectors}
Conformal killing vectors play a significant role in establishing the mathematical connection between the geometry of spacetime and the matter it contains through Einstein's field equations. These vectors are derived from the killing equations, utilizing the principles of Lie algebra \cite{waheed, chinese}. Conformal killing vectors are an essential tool for reducing the non-linearity order of field equations in various modified theories. In the context of general relativity, conformal killing vectors find numerous applications in geometric configurations, kinematics, and dynamics based on the structure theory. We employ an inheritance symmetry of spacetime characterized by conformal killing vectors, which are defined as \cite{arxiv,quantum} 
  
\begin{equation}\label{ckv}
	    \mathcal{L}_\eta g_{ij}=g_{k i}\eta^{k}_{;j}+g_{ik}\eta^k_{;j}=\zeta(r) g_{ij},
	\end{equation}
  where $\zeta, \eta^k$ and $g_{ij}$ represent the conformal factor, conformal killing vectors and metric tensor respectively. It is supposed that the vector $\eta$ generates the conformal symmetry and the metric $g$ is conformally mapped onto itself along $\eta$. The conformal factor, which characterizes the scaling of the metric, influence the geometry of the wormhole.
  By inserting the equation $\mathcal{L}_\eta g_{ij}=\zeta(r) g_{ij}$ from equation \eqref{ckv} into equation \eqref{whmetric}, we get the following equations:
  \begin{align}
       \eta^1 \Phi'(r)&=\dfrac{\zeta(r)}{2},\\
       \eta^1&=\dfrac{r \zeta(r)}{2},\\
       \eta^1 \left(-\dfrac{\Psi(r)-r\Psi'(r)}{r^2-r\Psi(r)}\right)+&2\eta^1_{,1}=\zeta(r).
  \end{align}
     
On solving the aforementioned expressions, we obtain the following two relationships for the metric components:
\begin{align}\label{crs}
    e^{2\Phi(r)}= \mathcal{C}_1 r^2, \quad and \quad\left(1-\frac{\Psi(r)}{r}\right)^{-1}= \dfrac{\mathcal{C}_2}{\zeta^2(r)},
\end{align}
  where, $\mathcal{C}_1$ and $\mathcal{C}_2$ are the integrating constants. For the simplification, we assume $A(r)=\zeta^2(r)$. Consequently, the expression for the shape function can be obtained as
  \begin{equation}\label{csf}
      \Psi(r)=r\left(1-\dfrac{A(r)}{\mathcal{C}_2}\right).
  \end{equation}
%%%%%%%%%%%%%%%%%%%%%%%%%%%%%%%%%%%%%%%%%%%%%%%%%%%%%%%%%%%%%%%%%%%%%%%%%%%
\section{Wormhole model in $\mathpzc{f}(\mathcal{Q},\mathcal{T})$ gravity}\label{section IV}
In this section, we shall consider a feasible model to study the properties of wormhole geometry. In particular, we suppose the linear form given by 
\begin{equation}
    \mathpzc{f}(\mathcal{Q},\mathcal{T})= \alpha \mathcal{Q}+\beta\mathcal{T},
\end{equation}
where $\alpha$ and $\beta$ are model parameter. For $\alpha=1, \beta=0$ one can retain General Relativity. By utilizing equations \eqref{crs}, \eqref{csf} and adopting dimensionless parameters, the field equations \eqref{fe1}-\eqref{fe3} can be solved to obtain the following equations: 
\begin{align}\label{nd1}
\rho_* \left(\dfrac{r}{\sqrt{\theta}}\right)&=\frac{\alpha  \left((\beta +6) \dfrac{r}{\sqrt{\theta}} \dot{A}_*\left(\dfrac{r}{\sqrt{\theta}}\right)+2 (8 \beta +3) A_*\left(\dfrac{r}{\sqrt{\theta}}\right)+2 (2 \beta -3) \mathcal{C}_2\right)}{6 (\beta +1) (2 \beta -1) \mathcal{C}_2 \dfrac{r^2}{\theta}},
\end{align}

\begin{align}\label{nd2}
   p_{r_*}\left(\dfrac{r}{\sqrt{\theta}}\right)&=-\frac{\alpha  \left(13 \beta  \dfrac{r}{\sqrt{\theta}}  \dot{A}_*\left(\dfrac{r}{\sqrt{\theta}}\right)+(18-8 \beta ) A_*\left(\dfrac{r}{\sqrt{\theta}}\right)+4 \beta  \mathcal{C}_2-6 \mathcal{C}_2\right)}{6 (\beta +1) (2 \beta -1)  \mathcal{C}_2 \dfrac{r^2}{\theta}},
\end{align}

\begin{align}\label{nd3}
p_{t_*}\left(\dfrac{r}{\sqrt{\theta}}\right)&=\frac{\alpha  \left(-(\beta +6) \dfrac{r}{\sqrt{\theta}}  \dot{A}_*\left(\dfrac{r}{\sqrt{\theta}}\right)-2 (8 \beta +3) A_*\left(\dfrac{r}{\sqrt{\theta}}\right)+8 \beta  \mathcal{C}_2\right)}{6 (\beta +1) (2 \beta -1)  \mathcal{C}_2 \dfrac{r^2}{\theta}}.
\end{align}
Here, the subscript '$\ast$' denotes corresponding adimensional quantities and the overhead dot is the derivative of the function with respect to $\frac{r}{\sqrt{\theta}}$. Further, non-dimensionalization is a powerful tool in theoretical physics. It enables researchers to simplify equations, comprehend the scaling behavior of physical systems, and gain insights into the essential features of complex phenomena such as wormholes.

Now, we shall discuss the physical analysis of wormhole solutions with the help of equations \eqref{nd1}-\eqref{nd3} under non-commutative distributions. For this purpose,  we consider the Gaussian and Lorentzian energy densities of the static and spherically symmetric particle-like gravitational source with a total mass of the form \cite{nicoli,math}
    \begin{equation}
         \label{gauss}\rho= \frac{M e^{-\frac{r^2}{4 \theta }}}{8 \pi ^{\frac{3}{2}} \theta ^{\frac{3}{2}}},
    \end{equation}
       and 
       \begin{equation}
            \label{lorentz}\rho=\frac{\sqrt{\theta } M}{\pi ^2 \left(\theta +r^2\right)^2}.
       \end{equation}.
%%%%%%%%%%%%%%%%%%%%%%%%%%%%%%%%%
\subsection{Gaussian energy density}
In this subsection, our attention will be directed towards exploring non-commutative geometry under Gaussian distribution. When we substitute the Gaussian energy density \eqref{gauss} into equation \eqref{nd1}, we obtain the resulting differential equation: 
\begin{equation}
    \frac{M_* e^{\left(-\frac{r^2}{4\theta}\right)}}{8 \pi ^{3/2}}=\frac{\alpha  \left((\beta +6) \dfrac{r}{\sqrt{\theta}}\dot{A}_*\left(\dfrac{r}{\sqrt{\theta}}\right)+2 (8 \beta +3) A_*\left(\dfrac{r}{\sqrt{\theta}}\right)+2 (2 \beta -3)\mathcal{C}_2\right)}{6 (\beta +1) (2 \beta -1)\mathcal{C}_2 \dfrac{r^2}{\theta}},
\end{equation}
where $M_*$ has the dimension of $\frac{r}{\sqrt{\theta}}$ and can be expressed as $\frac{M}{\sqrt{\theta}}$. An observation is that the shape function $\Psi_*$ to fulfill the throat condition, we enforce the initial condition $A_*\left(\frac{r_0}{\sqrt{\theta}}\right)=0$ utilizing the relation \eqref{csf}. Then we have the particular solution,
\begin{equation}
\begin{split}
    A_*\left(\dfrac{r}{\sqrt{\theta}}\right)=\dfrac{\mathcal{C}_2 \left(\dfrac{r}{\sqrt{\theta}}\right)^{\frac{90}{\beta +6}-16}}{\pi ^{3/2} \alpha  (\beta +6) (8 \beta +3)}\Bigg[&-\frac{3}{8} (\beta +1) (2 \beta -1) (8 \beta +3) M_*\Bigg.\\
    &\Bigg.
    \left(\left(\dfrac{r}{\sqrt{\theta}}\right)^{\frac{18 (\beta +1)}{\beta +6}} E_{\frac{45}{\beta +6}-8}\left(\frac{r^2}{4\theta}\right)-\left(\dfrac{r_0}{\sqrt{\theta}}\right)^{\frac{18 (\beta +1)}{\beta +6}} E_{\frac{45}{\beta +6}-8}\left(\frac{r_0^2}{4\theta}\right)\right)\Bigg.\\
    &\Bigg.
    -\pi ^{3/2} \alpha  (\beta +6) (2 \beta -3) \left(\left(\dfrac{r}{\sqrt{\theta}}\right)^{\frac{15 \beta }{\beta +6}+1}-\left(\dfrac{r_0}{\sqrt{\theta}}\right)^{\frac{15 \beta }{\beta +6}+1}\right)\Bigg].
    \end{split}
\end{equation}
Here, $E$ is a special function defined as $E_n(x)=-\int_{-x}^{\infty} \frac{e^{-t}}{t^n}dt$. Consequently, the corresponding shape function obtained as,
\begin{equation}\label{nsf}
\begin{split}
    \Psi_*\left(\dfrac{r}{\sqrt{\theta}}\right)=&\frac{3 \left(2 \beta ^2+\beta -1\right) M_* \left(\dfrac{r}{\sqrt{\theta}}\right)^{-\frac{15 \beta }{\beta +6}} \left(\left(\dfrac{r}{\sqrt{\theta}}\right)^{\frac{18 (\beta +1)}{\beta +6}} E_{\frac{45}{\beta +6}-8}\left(\frac{r^2}{4\theta}\right)-\left(\dfrac{r_0}{\sqrt{\theta}}\right)^{\frac{18 (\beta +1)}{\beta +6}} E_{\frac{45}{\beta +6}-8}\left(\frac{r_0^2}{4\theta}\right)\right)}{8 \pi ^{3/2} \alpha  (\beta +6)}\\
    &+\frac{(3-2 \beta ) \left(\dfrac{r}{\sqrt{\theta}}\right)^{-\frac{15 \beta }{\beta +6}} \left(\dfrac{r_0}{\sqrt{\theta}}\right)^{\frac{15 \beta }{\beta +6}+1}+10 \beta \left(\dfrac{r}{\sqrt{\theta}}\right)}{8 \beta +3}.
    \end{split}
\end{equation}

\begin{figure*}[!]
	    \subfloat[$\Psi_*\left(\frac{r}{\sqrt{\theta}}\right) >0$ \label{fig:gsf1}]{\includegraphics[width=0.4\linewidth]{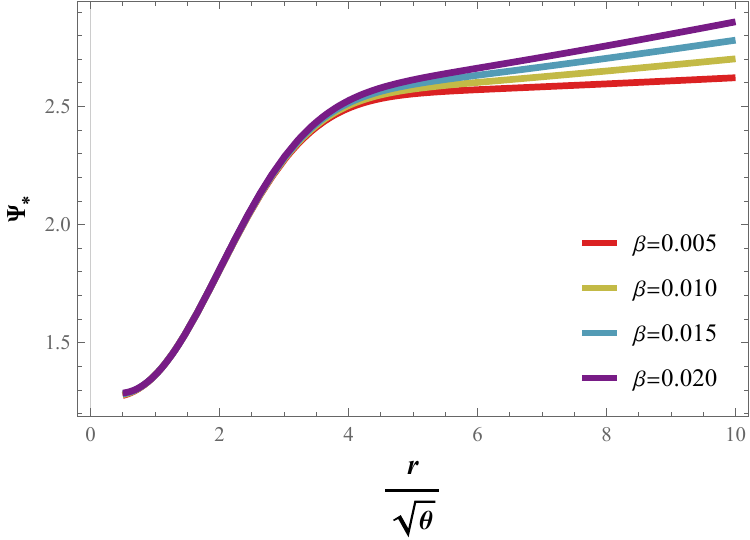}}
      \subfloat[$\dfrac{\Psi_*\left(\frac{r}{\sqrt{\theta}}\right)-\frac{r}{\sqrt{\theta}}\dot{\Psi}_*\left(\frac{r}{\sqrt{\theta}}\right)}{\Psi_*\left(\frac{r}{\sqrt{\theta}}\right)^2}>0$\label{fig:gsf4}]{\includegraphics[width=0.445\linewidth]{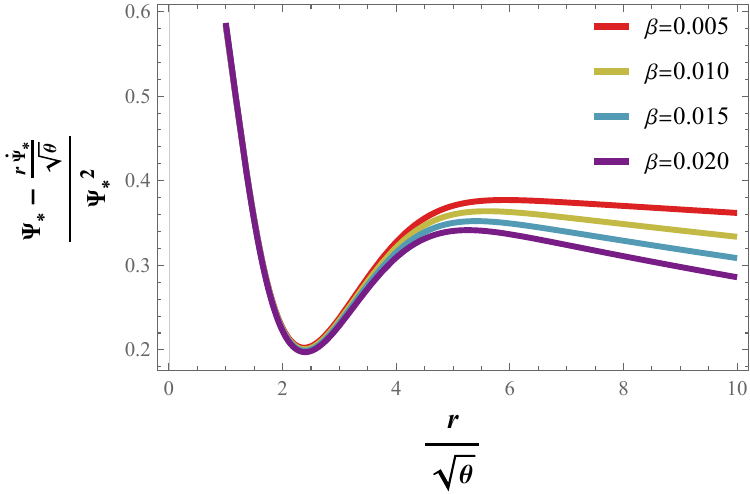}}\\
      \subfloat[$\dot{\Psi}_*\left(\frac{r}{\sqrt{\theta}}\right)<1$\label{fig:gsf2}]{\includegraphics[width=0.4\linewidth]{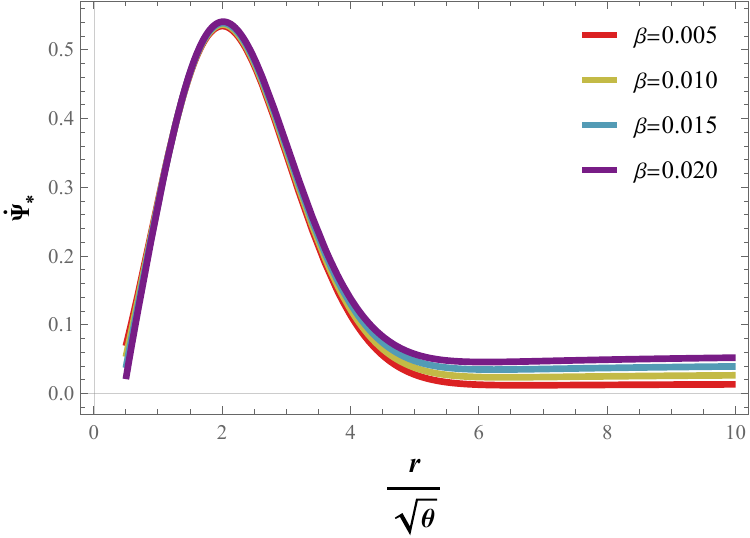}}
       \subfloat[$\dfrac{\Psi_*\left(\frac{r}{\sqrt{\theta}}\right)}{\frac{r}{\sqrt{\theta}}}\to 0$ as $\frac{r}{\sqrt{\theta}}\to \infty$\label{fig:gsf3}]{\includegraphics[width=0.43\linewidth]{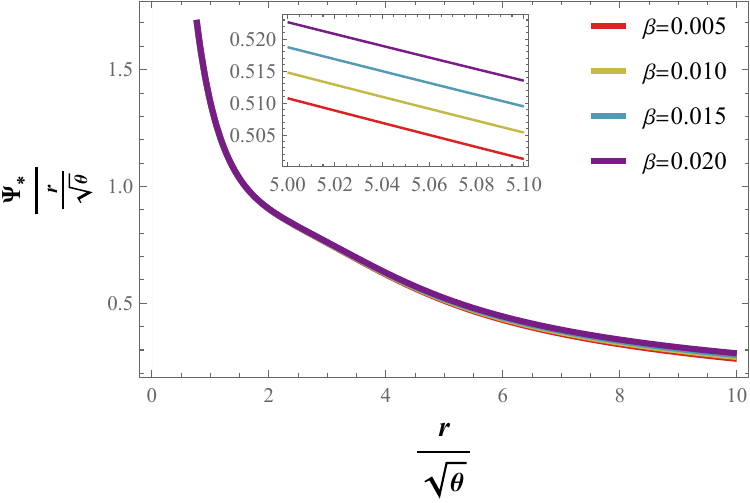}}
	    \caption{The graphical behavior of shape function $\Psi_*$ for Gaussian non-commutative geometry with $M_*=7.25, \alpha=0.45, \mathcal{C}_2=2$ and $\frac{r_0}{\sqrt{\theta}}=1.6$.}
	    \label{fig:gsf}
	   \end{figure*}

 \begin{figure*}[b!]
	    \centering
	    \subfloat[Energy density $\rho_*$\label{fig:grho}]{\includegraphics[width=0.4\linewidth]{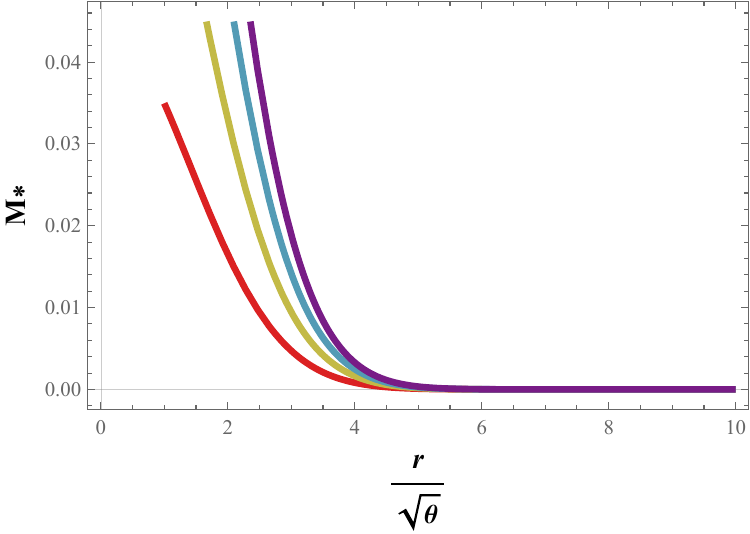}}
	    \subfloat[NEC $\rho_*+p_{r_*}$\label{fig:ge1}]{\includegraphics[width=0.405\linewidth]{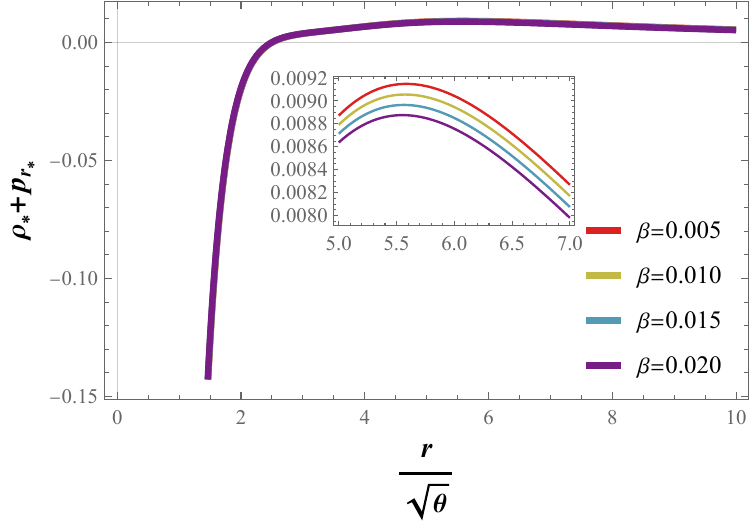}}\\
	    \subfloat[NEC $\rho_*+p_{t_*}$\label{fig:ge2}]{\includegraphics[width=0.4\linewidth]{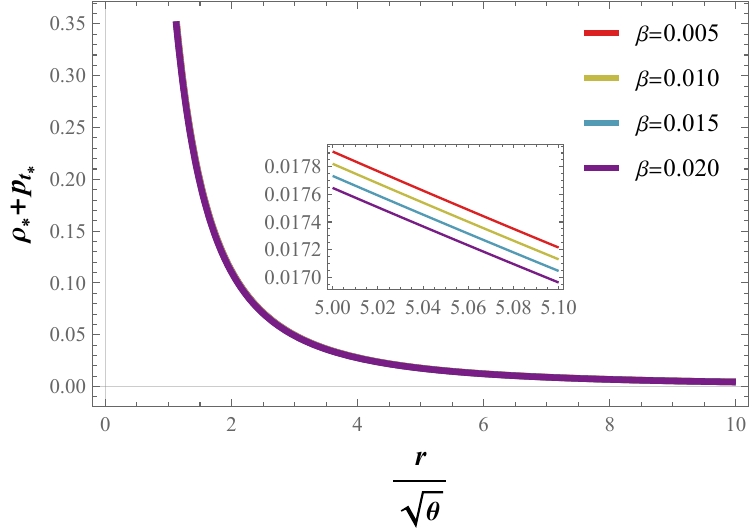}}
	    \subfloat[DEC $\rho_*-|p_{r_*}|$\label{fig:ge3}]{\includegraphics[width=0.405\linewidth]{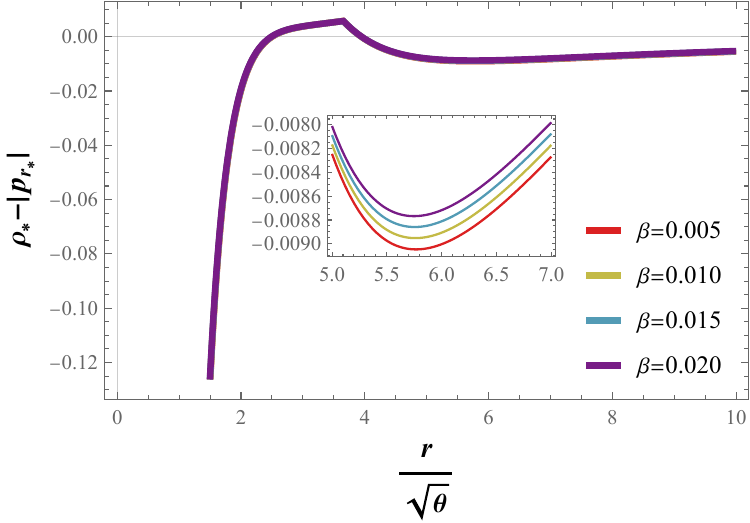}}\\
	    \subfloat[DEC $\rho_*-|p_{t_*}|$\label{fig:ge4}]{\includegraphics[width=0.4\linewidth]{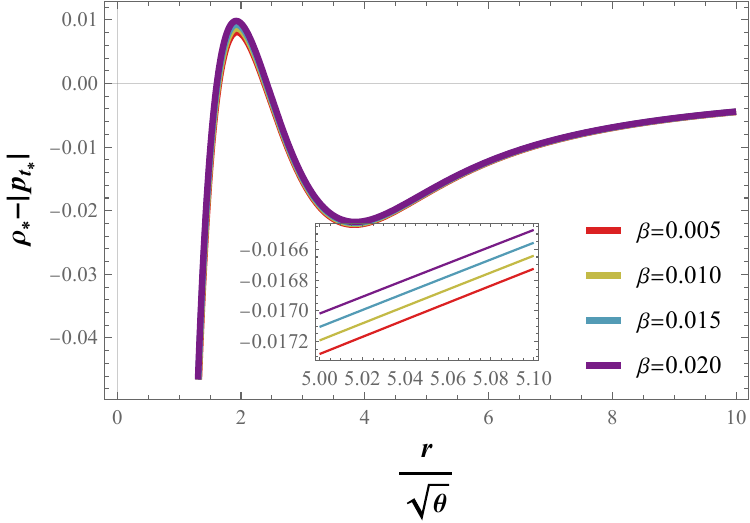}}
	    \subfloat[SEC $\rho_*+p_{r_*}+2p_{t_*}$\label{fig:ge5}]{\includegraphics[width=0.405\linewidth]{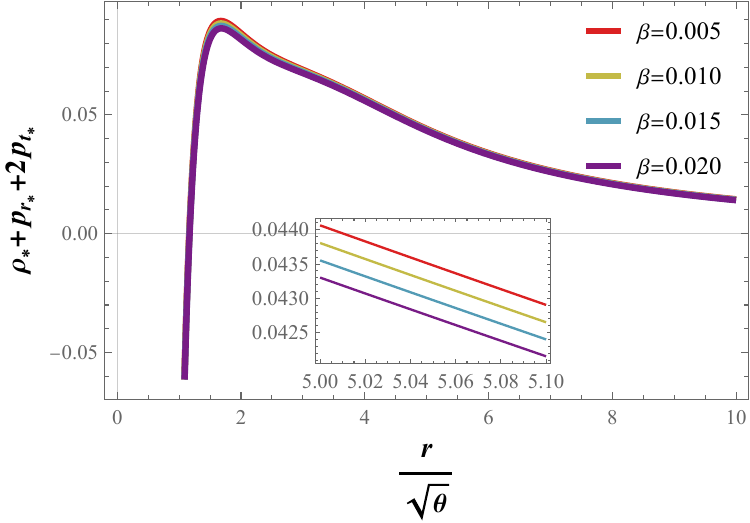}}
	    \caption{ Gaussian Source: The profile of energy density and energy conditions with respect to $\frac{r}{\sqrt{\theta}}$ for different values of $\beta$ with fixed parameters $M_*=7.25, \alpha=0.45, \mathcal{C}_2=2$ and $\frac{r_0}{\sqrt{\theta}}=1.6$. }
	    \label{fig:gec}
	\end{figure*}
We can easily verify the satisfaction of the throat condition by performing a simple calculation of $\Psi_*\left(\frac{r_0}{\sqrt{\theta}}\right)$. Furthermore, by evaluating the derivative of the shape function \eqref{nsf} at the throat, we derive the following relation:   
\begin{equation}
    \dot{\Psi}_*\left(\dfrac{r_0}{\sqrt{\theta}}\right)=\frac{5 \beta }{\beta +6}+\frac{3 (\beta +1) (2 \beta -1) M_* \dfrac{r_0^2}{\theta} \left(36 (\beta +1) E_{\frac{45}{\beta +6}-8}\left(\frac{r_0^2}{4\theta}\right)-(\beta +6) \dfrac{r_0^2}{\theta} E_{-\frac{9 (\beta +1)}{\beta +6}}\left(\frac{r_0^2}{4\theta}\right)\right)}{16 \pi ^{3/2} \alpha  (\beta +6)^2}.
\end{equation}

 For the present scenario,  we explain some properties of the obtained shape function in the background of Gaussian distribution. It can be seen that the shape function $ \Psi_*\left(\dfrac{r}{\sqrt{\theta}}\right)$ depends on the non-commutative parameters and the coupling parameters. Initially, we select the different values of $\beta \in [0,0.5)$ and analyze the results depending on the choice of other parameters. In this case, we pick the suitable values of parameters $M_*=7.25, \alpha=0.45$ and $\mathcal{C}_2=2$. Further, the throat of the wormhole is located at  $\frac{r_0}{\sqrt{\theta}}=1.6$. Figure \ref{fig:gsf} shows the characteristics of $ \Psi_*\left(\dfrac{r}{\sqrt{\theta}}\right)$ with different values of coupling constant $\beta$. Clearly, the obtained shape function $ \Psi_*\left(\dfrac{r}{\sqrt{\theta}}\right)>0$ is a monotonically increasing function $\forall \,\,\frac{r}{\sqrt{\theta}}> \frac{r_0}{\sqrt{\theta}}$ [Figure \ref{fig:gsf1}]. From Figures \ref{fig:gsf4}, \ref{fig:gsf2} indicates that $\dfrac{\Psi_*\left(\frac{r}{\sqrt{\theta}}\right)-\frac{r}{\sqrt{\theta}}\dot{\Psi}_*\left(\frac{r}{\sqrt{\theta}}\right)}{\Psi_*\left(\frac{r}{\sqrt{\theta}}\right)^2}>0$ and the derivative of shape function is less than one, which confirms that the flaring-out condition is satisfied for a linear model. The satisfaction of the asymptotic condition for our model is depicted in Figure \ref{fig:gsf3}, when the ratio $\dfrac{\Psi_*\left(\frac{r}{\sqrt{\theta}}\right)}{\frac{r}{\sqrt{\theta}}}\to 0$ for the value of the radial coordinates $\frac{r}{\sqrt{\theta}}$ is extremely large. 
 
Now, by substituting the $A_*$ function into \eqref{nd2} and \eqref{nd3}, we get the following pressure elements:
\begin{gather}
\begin{split}
    p_{r_*}=-\frac{\alpha} {32 (\beta +1) (\beta +6)^2 \left(\frac{r^{18}}{\theta^9}\right)} &\Bigg[\frac{(\beta +1) M_* }{\pi ^{3/2} \alpha }\Biggl(13 \beta  (\beta +6) \left(\frac{r^{20}}{\theta^{10}}\right) E_{-\frac{9 (\beta +1)}{\beta +6}}\left(\frac{r^2}{4\theta}\right)\Biggl.\Bigg.\\
    \Biggl.\Bigg.
    &+36 (\beta +1)
    \Biggl(-6 (2 \beta -1) \left(\dfrac{r}{\sqrt{\theta}}\right)^{\frac{90}{\beta +6}} \left(\dfrac{r_0}{\sqrt{\theta}}\right)^{\frac{18 (\beta +1)}{\beta +6}} E_{\frac{45}{\beta +6}-8}\left(\frac{r_0^2}{4\theta}\right)\Biggl.\Biggl.\Bigg.\\
    \Biggl.\Biggl.\Bigg.
   & -(\beta +6) \left(\dfrac{r^{18}}{\theta^9}\right) E_{\frac{45}{\beta +6}-8}\left(\frac{r^2}{4\theta}\right)\Biggl)\Biggl) \Bigg.\\
    \Bigg.
    &+\frac{64 (\beta +6) (2 \beta -3) \left((\beta +6)\left(\dfrac{r^{16}}{\theta^8}\right)-9 (\beta +1) \left(\dfrac{r}{\sqrt{\theta}}\right)^{\frac{90}{\beta +6}}\left(\dfrac{r_0}{\sqrt{\theta}}\right)^{\frac{15 \beta }{\beta +6}+1}\right)}{8 \beta +3}\Bigg],
\end{split}
\end{gather}

\begin{equation}
     p_{t_*}=\frac{M_* \left(36 (\beta +1) E_{\frac{45}{\beta +6}-8}\left(\frac{r^2}{4\theta}\right)-(\beta +6)\left(\frac{r^2}{4\theta}\right) E_{-\frac{9 (\beta +1)}{\beta +6}}\left(\frac{r^2}{4\theta}\right)\right)}{32 \pi ^{3/2} (\beta +6)}+\frac{\alpha }{(\beta +1) \left(\frac{r^2}{\theta}\right)}.
\end{equation}

In our study, the behavior of energy density and energy conditions are illustrated in Figure \ref{fig:gec}. Both dominant energy conditions, radial null energy condition and strong energy condition are violated. However, the tangential null energy condition is satisfied.

%%%%%%%%%%%%%%%%%%%%%%%%%%%%%%%%%%%%%%%%%%%%%%%%%%%%%%%%%%%%%%%%%%%%%
\subsection{Lorentzian energy density}
In this subsection, we focus on the scenario involving non-commutative geometry with the Lorentzian distribution. By substituting the Lorentzian energy density \eqref{lorentz} into \eqref{nd1}, we get
\begin{equation}
    \frac{M_*}{\pi ^2 \left(\dfrac{r^2}{\theta}+1\right)^2}=\frac{\alpha  \left((\beta +6) \left(\dfrac{r}{\sqrt{\theta}}\right) \dot{A}_*\left(\dfrac{r}{\sqrt{\theta}}\right)+2 (8 \beta +3) A_*\left(\dfrac{r}{\sqrt{\theta}}\right)+2 (2 \beta -3) \mathcal{C}_2\right)}{6 (\beta +1) (2 \beta -1) \mathcal{C}_2 \dfrac{r^2}{\theta}}.
\end{equation}

Solving the aforementioned differential equation while imposing the throat condition on the shape function, we can derive the following expression:
\begin{equation}\label{nf}
\begin{split}
     A_*\left(\dfrac{r}{\sqrt{\theta}}\right)=&\dfrac{\mathcal{C}_2 \left(\dfrac{r}{\sqrt{\theta}}\right)^{\frac{90}{\beta +6}-16}} {\pi ^2 \alpha  (8 \beta +3)}\Bigg[3 \left(2 \beta ^2+\beta -1\right) M_* \Biggl(\left(\dfrac{r}{\sqrt{\theta}}\right)^{\frac{15 \beta }{\beta +6}+1} \Bigg(\, _2F_1\left(1,\frac{8 \beta +3}{\beta +6};\frac{9 (\beta +1)}{\beta +6};-\dfrac{r^2}{\theta}\right)\Bigg.\Biggl.\Bigg.\\
     \Bigg.\Biggl.\Bigg.&
     -\, _2F_1\left(2,\frac{8 \beta +3}{\beta +6};\frac{9 (\beta +1)}{\beta +6};-\dfrac{r^2}{\theta}\right)\Biggl)+\left(\dfrac{r_0}{\sqrt{\theta}}\right)^{\frac{15 \beta }{\beta +6}+1} \Bigg(\, _2F_1\left(2,\frac{8 \beta +3}{\beta +6};\frac{9 (\beta +1)}{\beta +6};-\dfrac{r_0^2}{\theta}\right)\Bigg.\Biggl.\Bigg.\\
       \Bigg.\Biggl.\Bigg.&
     -\, _2F_1\left(1,\frac{8 \beta +3}{\beta +6};\frac{9 (\beta +1)}{\beta +6};-\dfrac{r_0^2}{\theta}\right)\Bigg)\Biggl)-\pi ^2 \alpha  (2 \beta -3) \left(\left(\dfrac{r}{\sqrt{\theta}}\right)^{\frac{15 \beta }{\beta +6}+1}-\left(\dfrac{r_0}{\sqrt{\theta}}\right)^{\frac{15 \beta }{\beta +6}+1}\right)\Bigg],
\end{split}
\end{equation}

 where $ _2F_1( a, b;c;z)$ is the hypergeometric function. Hence, the resulting shape function can be expressed as follows:
\begin{equation}
\begin{split}
    \Psi_*\left(\dfrac{r}{\sqrt{\theta}}\right)=&\dfrac{1}{8 \beta +3}\Bigg[\dfrac{3 \left(2 \beta ^2+\beta -1\right)M_* \left(\dfrac{r}{\sqrt{\theta}}\right)^{-\frac{15 \beta }{\beta +6}}} {\pi ^2 \alpha }\Bigg(\left(\dfrac{r}{\sqrt{\theta}}\right)^{\frac{15 \beta }{\beta +6}+1} \Biggl(\, _2F_1\left(2,\frac{8 \beta +3}{\beta +6};\frac{9 (\beta +1)}{\beta +6};-\dfrac{r^2}{\theta}\right)\Biggl.\Bigg.\Bigg.\\
    \Biggl.\Bigg.\Bigg.&
    -\, _2F_1\left(1,\frac{8 \beta +3}{\beta +6};\frac{9 (\beta +1)}{\beta +6};-\dfrac{r^2}{\theta}\right)\Biggl)+\left(\dfrac{r_0}{\sqrt{\theta}}\right)^{\frac{15 \beta }{\beta +6}+1} \Biggl(\, _2F_1\left(1,\frac{8 \beta +3}{\beta +6};\frac{9 (\beta +1)}{\beta +6};-\dfrac{r_0^2}{\theta}\right)\Biggl.\Bigg.\Bigg.\\
    \Biggl.\Bigg.\Bigg.&
    -\, _2F_1\left(2,\frac{8 \beta +3}{\beta +6};\frac{9 (\beta +1)}{\beta +6};-\dfrac{r_0^2}{\theta}\right)\Biggl)\Bigg)+(3-2 \beta ) \left(\dfrac{r}{\sqrt{\theta}}\right)^{-\frac{15 \beta }{\beta +6}} \left(\dfrac{r_0}{\sqrt{\theta}}\right)^{\frac{15 \beta }{\beta +6}+1}+10 \beta  \dfrac{r}{\sqrt{\theta}}\Bigg],
    \end{split}
\end{equation}

\begin{figure*}[!]
	    \subfloat[$\Psi_*\left(\frac{r}{\sqrt{\theta}}\right) >0$ \label{fig:lsf1}]{\includegraphics[width=0.4\linewidth]{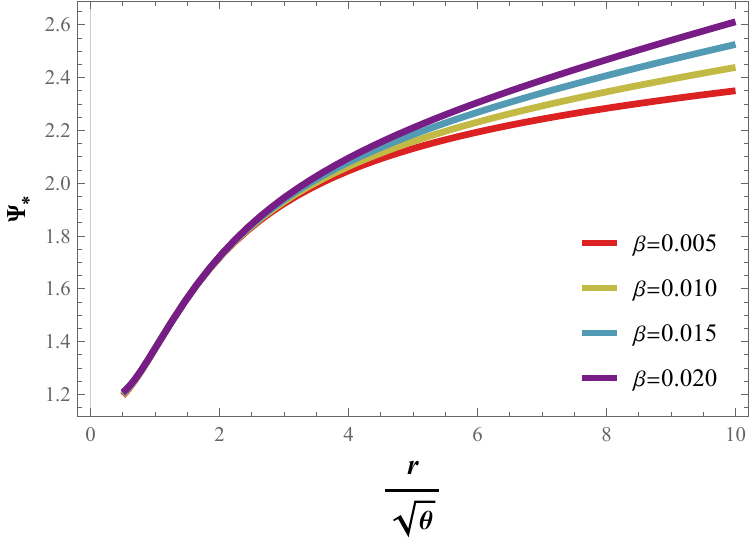}}
	  \subfloat[$\dfrac{\Psi_*\left(\frac{r}{\sqrt{\theta}}\right)-\frac{r}{\sqrt{\theta}}\dot{\Psi}_*\left(\frac{r}{\sqrt{\theta}}\right)}{\Psi_*\left(\frac{r}{\sqrt{\theta}}\right)^2}>0$\label{fig:lsf4}]{\includegraphics[width=0.44\linewidth]{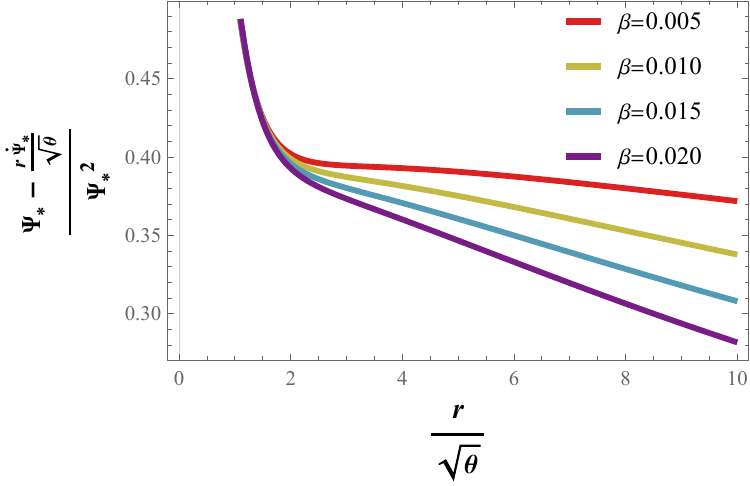}} \\
      \subfloat[$\dot{\Psi}_*\left(\frac{r}{\sqrt{\theta}}\right)<1$\label{fig:lsf2}]{\includegraphics[width=0.4\linewidth]{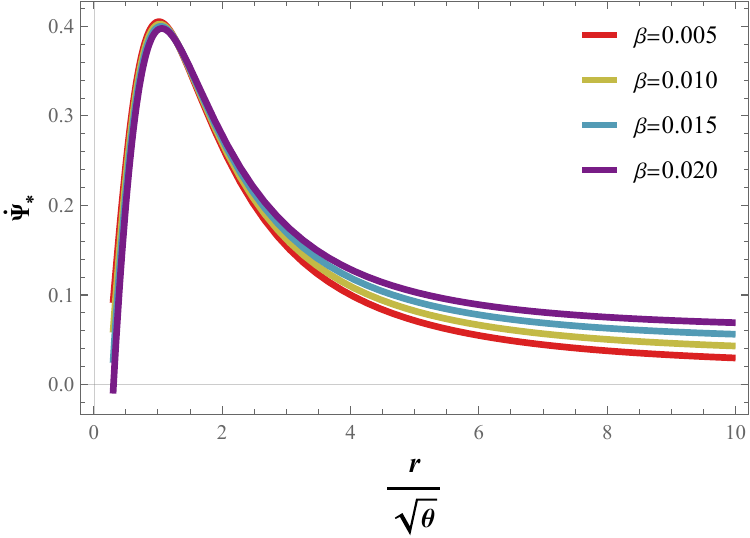}} 
      \subfloat[$\dfrac{\Psi_*\left(\frac{r}{\sqrt{\theta}}\right)}{\frac{r}{\sqrt{\theta}}}\to 0$ as $\frac{r}{\sqrt{\theta}}\to \infty$\label{fig:lsf3}]{\includegraphics[width=0.43\linewidth]{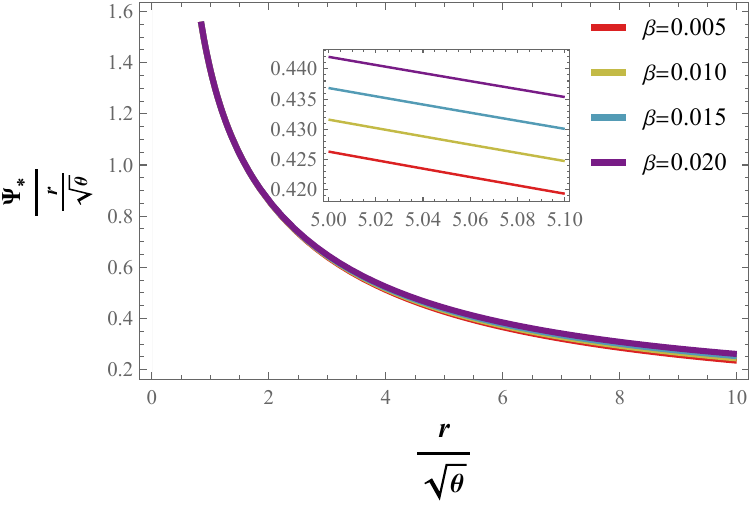}}
	    \caption{The graphical behavior of shape function $\Psi_*$ for Lorentzian non-commutative geometry with $M_*=7.25, \alpha=0.45, \mathcal{C}_2=2$ and $\frac{r_0}{\sqrt{\theta}}=1.6$.}
	    \label{fig:lsf}
	   \end{figure*}

\begin{figure*}[b!]
	    \centering
	    \subfloat[Energy density $\rho_*$\label{fig:lrho}]{\includegraphics[width=0.41\linewidth]{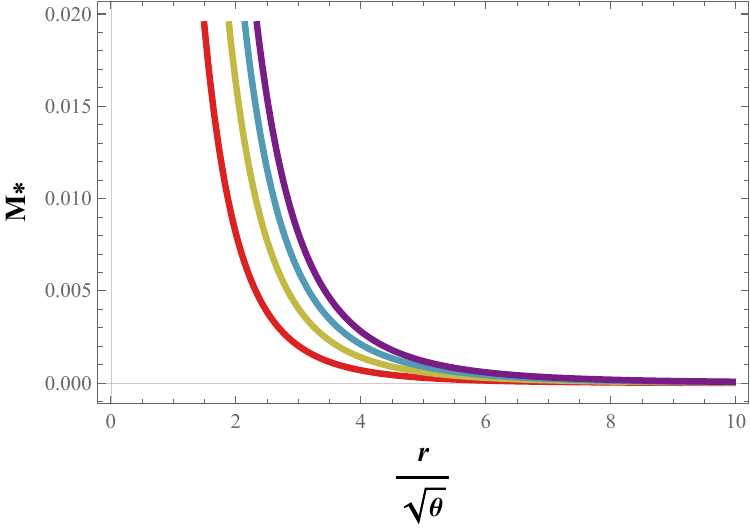}}
	    \subfloat[NEC $\rho_*+p_{r_*}$\label{fig:le1}]{\includegraphics[width=0.4\linewidth]{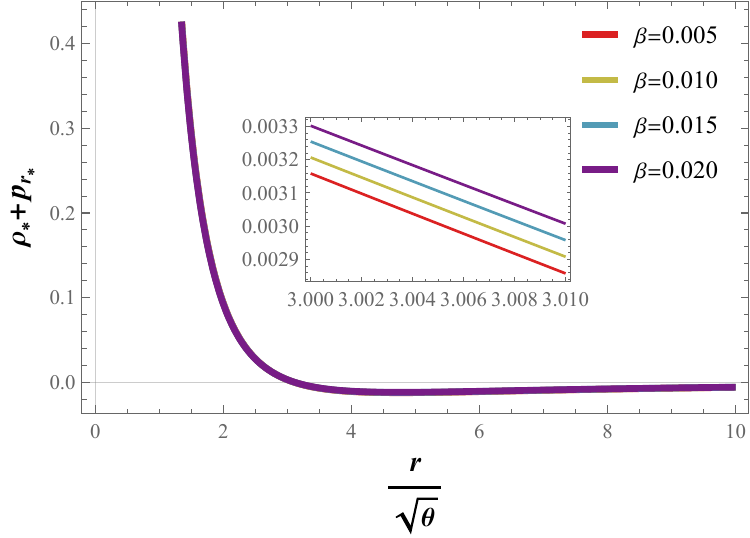}}\\
	    \subfloat[NEC $\rho_*+p_{t_*}$\label{fig:le2}]{\includegraphics[width=0.4\linewidth]{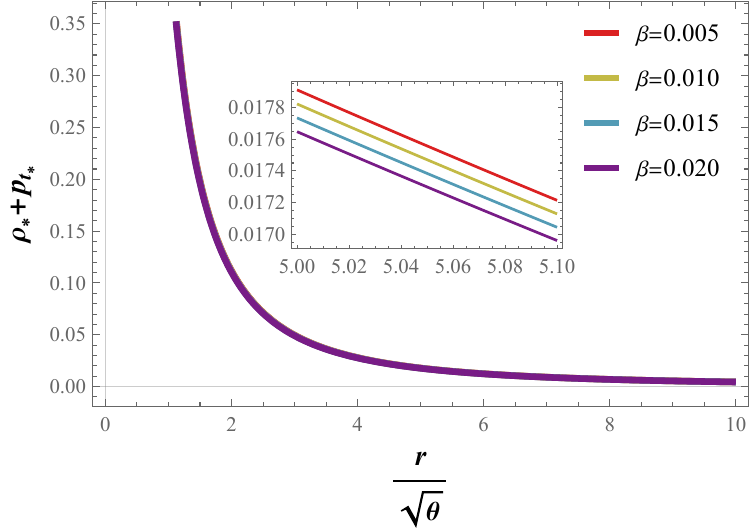}}
	    \subfloat[DEC $\rho_*-|p_{r_*}|$\label{fig:le3}]{\includegraphics[width=0.4052\linewidth]{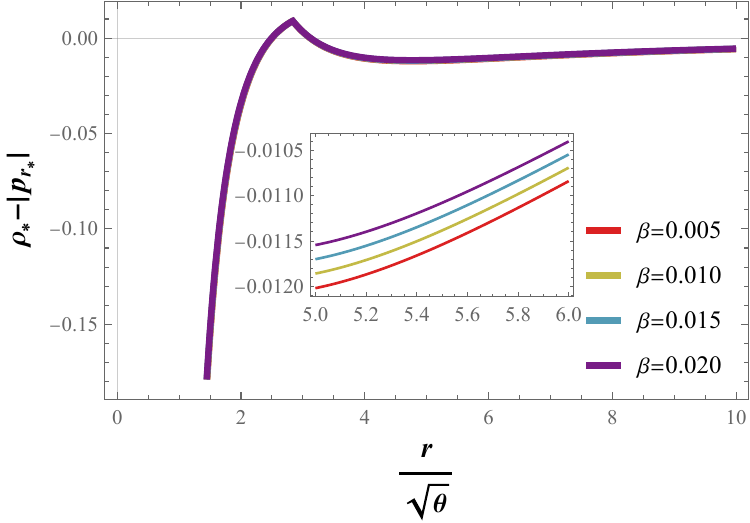}}\\
	    \subfloat[DEC $\rho_*-|p_{t_*}|$\label{fig:le4}]{\includegraphics[width=0.41\linewidth]{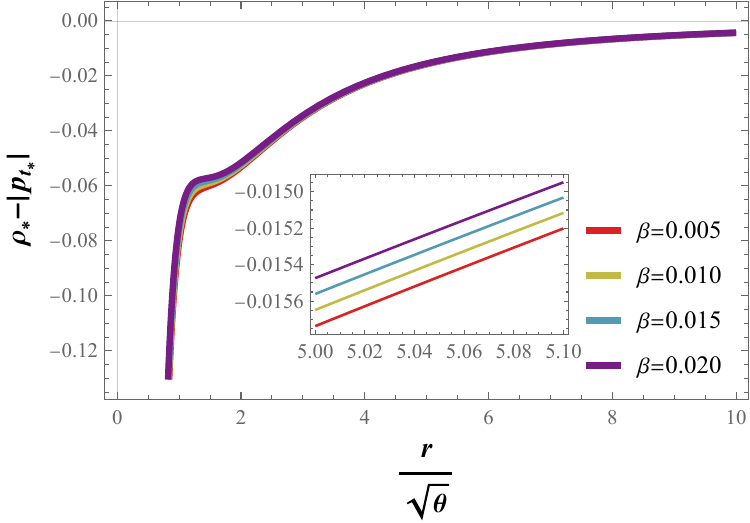}}
	    \subfloat[SEC $\rho_*+p_{r_*}+2p_{t_*}$\label{fig:le5}]{\includegraphics[width=0.4\linewidth]{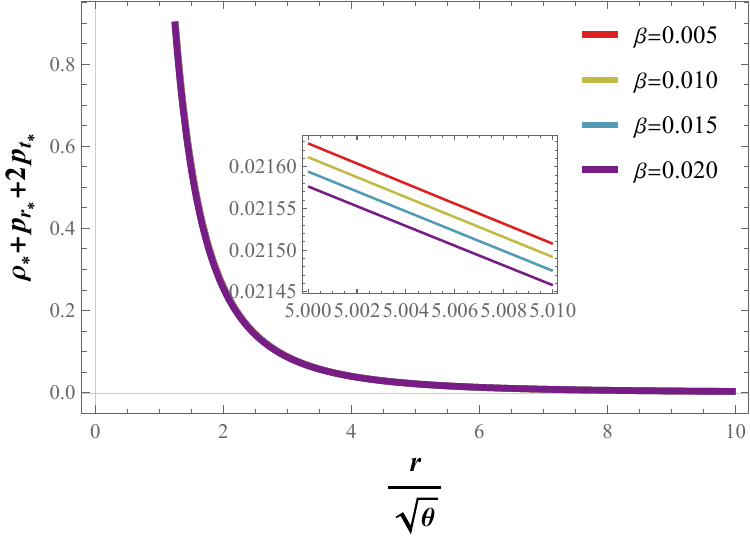}}
	    \caption{Lorentzian Source: The profile of energy density and energy conditions with respect to $\frac{r}{\sqrt{\theta}}$ for different values of $\beta$ with fixed parameters $M_*=7.25, \alpha=0.45, \mathcal{C}_2=2$ and $\frac{r_0}{\sqrt{\theta}}=1.6$.}
	    \label{fig:lec}
	\end{figure*}
From the above expression, the derivative of the shape function is given by  
\begin{equation}
     \dot{\Psi}_*\left(\dfrac{r_0}{\sqrt{\theta}}\right)=\frac{5 \beta -\frac{6 \left(2 \beta ^2+\beta -1\right) M_* \dfrac{r_0^2}{\theta}}{\pi ^2 \alpha  \left(\dfrac{r_0^2}{\theta}+1\right)^2}}{\beta +6}.
\end{equation}

Here, we will present the graphical representation of the obtained shape function as well as its criteria that must be fulfilled for the choice of the parameters $M_*=7.25, \alpha=0.4, \mathcal{C}_2=2$ and $\frac{r_0}{\sqrt{\theta}}=1.6$ to analyze the results fruitfully. From Figure \ref{fig:lsf1}, the positive and increased nature of shape function $ \Psi_*\left(\dfrac{r}{\sqrt{\theta}}\right)$ reveals that the Lorentzian distribution is satisfied the wormhole geometry for conformal symmetry in the context of $\mathpzc{f}(\mathcal{Q},\mathcal{T})$ gravity. Furthermore, the obtained shape function satisfies the flaring-out condition in Figure \ref{fig:lsf4}, \ref{fig:lsf2}. For an infinitely large value of the radial coordinate $\dfrac{\Psi_*\left(\frac{r}{\sqrt{\theta}}\right)}{\frac{r}{\sqrt{\theta}}}\to 0$ [Figure \ref{fig:lsf3}]. Therefore, we can assert that the obtained shape function fulfills all the necessary requirements of the wormhole.

Now, substituting function \eqref{nf} into \eqref{nd2} and \eqref{nd3}, we get the pressure elements as
\begin{equation}
\begin{split}
    p_{r_*}=&\dfrac{1}{\left(\dfrac{r^{18}}{\theta^9}\right)}\Bigg[\dfrac{6 (2 \beta -1) M_* \Gamma \left(\frac{5 (2 \beta +3)}{\beta +6}\right)} {\pi ^2 (8 \beta +3)}\Biggl(\left(\dfrac{r}{\sqrt{\theta}}\right)^{\frac{90}{\beta +6}} \left(\dfrac{r_0}{\sqrt{\theta}}\right)^{\frac{15 \beta }{\beta +6}+1} \Bigg(\, _2\tilde{F}_1\left(2,\frac{8 \beta +3}{\beta +6};\frac{9 (\beta +1)}{\beta +6};-\dfrac{r_0^2}{\theta}\right)\Bigg.\Biggl.\Bigg.\\
    \Bigg.\Biggl.\Bigg.&
    -\, _2\tilde{F}_1\left(1,\frac{8 \beta +3}{\beta +6};\frac{9 (\beta +1)}{\beta +6};-\dfrac{r_0^2}{\theta}\right)\Bigg)
    +\dfrac{r^{16}}{\theta^8} \Bigg(\, _2\tilde{F}_1\left(1,\frac{8 \beta +3}{\beta +6};\frac{9 (\beta +1)}{\beta +6};-\dfrac{r^2}{\theta}\right)\Bigg.\Biggl.\Bigg.\\
    \Bigg.\Biggl.\Bigg.&
    -\, _2\tilde{F}_1\left(2,\frac{8 \beta +3}{\beta +6};\frac{9 (\beta +1)}{\beta +6};-\dfrac{r^2}{\theta}\right)\Bigg)\Biggl)\Bigg.\\
    \Bigg.&
    +\dfrac{1}{\beta +6}\Biggl(\dfrac{2 \alpha  (2 \beta -3) \left(9 (\beta +1) \left(\dfrac{r}{\sqrt{\theta}}\right)^{\frac{90}{\beta +6}} \left(\dfrac{r_0}{\sqrt{\theta}}\right)^{\frac{15 \beta }{\beta +6}+1}-(\beta +6) \dfrac{r^{16}}{\theta^8}\right)}{(\beta +1) (8 \beta +3)}
    -\dfrac{13 \beta  M_* \dfrac{r^{18}}{\theta^9}}{\pi ^2 \left(x^2+1\right)^2}\Biggl)\Bigg],
    \end{split}
\end{equation}

\begin{equation}
    p_{t_*}=\frac{\alpha }{(\beta +1) \dfrac{r^2}{\theta}}-\frac{M_*}{\pi ^2 \left(\dfrac{r^2}{\theta}+1\right)^2}.
\end{equation}
 where $\Gamma(a, z)$ is the gamma function.

 Figure \ref{fig:lec} illustrates characteristics of the energy conditions and the corresponding energy density profile for Lorentzian distribution. It shows that in this scenario, the radial null energy condition [Figure \ref{fig:le1}] and dominant energy conditions [Figure \ref{fig:le3}, \ref{fig:le3}] are violated. But, the tangential null energy condition [Figure \ref{fig:le2}] and strong energy condition [Figure \ref{fig:le5}] are obeyed.

Moreover, by investigating the existence of wormhole solutions and analyzing energy conditions in the late-time universe, we explore exotic matter and energy distributions that could enable the formation and stability of wormholes. The presence or absence of these solutions has significant implications for our understanding of the late-time universe's evolution and the nature of exotic matter needed to support such structures.
%%%%%%%%%%%%%%%%%%%%%%%%%%%%%%%%%%%%%%%%%%%%%%%%%%%%%%%%%%%%%%%%%%%%%%%%%%%
\section{Effect of Anisotropy}\label{section V}
In this section, we explore the anisotropy of Gaussian and Lorentzian non-commutative geometry in order to understand the characteristics of the anisotropic pressure. The quantification of anisotropy plays a crucial role in revealing the internal geometry of a relativistic wormhole configuration. It is well known that the level of anisotropy within a wormhole can be measured using the following formula \cite{mustafa, zinnat, kalam, waseem, shamir, shamir1}:
\begin{equation}
\Delta= p_{t_*}-p_{r_*}.
\end{equation}

\begin{figure}[b!]
    \centering
   \subfloat[Gaussian distribution: $p_{t_*}-p_{r_*}>0$\label{fig:gaa}]{\includegraphics[width=0.465\linewidth]{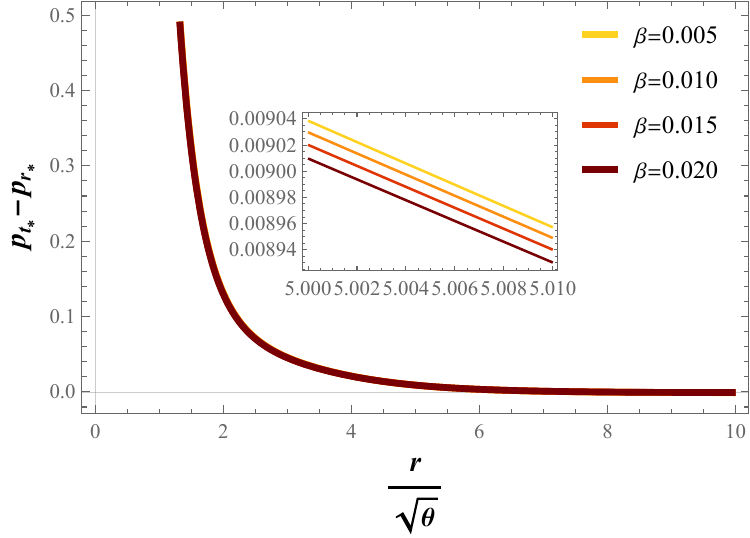}}
	    \subfloat[Lorentzian distribution: $p_{t_*}-p_{r_*}<0$\label{fig:laa}]{\includegraphics[width=0.48\linewidth]{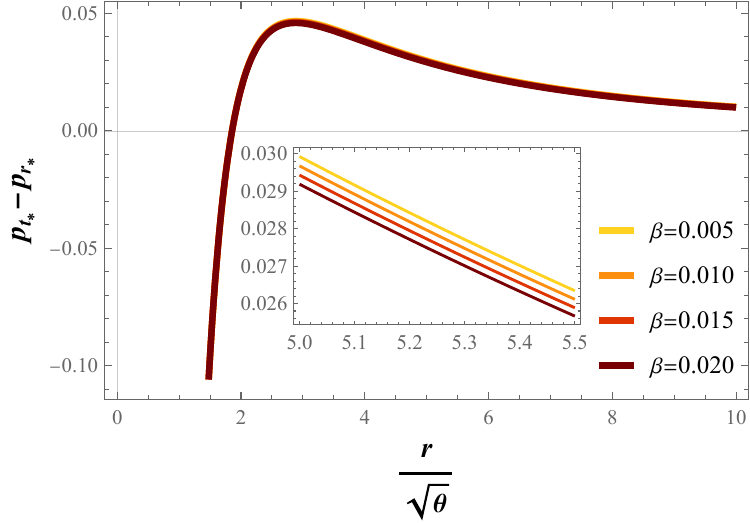}}
    \caption{The graphical representation of anisotropy for both distributions.}
    \label{fig:anisotropy}
\end{figure}

We can determine the geometry of the wormhole based on anisotropic factor. When the tangential pressure is greater than the radial pressure, it results in $\Delta>0$. This signifies that the structure of the wormhole is repulsive and anisotropic force is acting outward direction. Conversely, if the radial pressure is greater than the tangential pressure, it yields $\Delta<0$. This indicates an attractive geometry of the wormhole and force is directed inward. The anisotropy for both the Gaussian ($\Delta_G$) and Lorentzian ($\Delta_L$) distributions with
the linear model is calculated as
\begin{equation}
\begin{split}
    \Delta_G=& 
    \frac{3 \left(\dfrac{r}{\sqrt{\theta}}\right)^{-\frac{18 (\beta +1)}{\beta +6}}} {4 (\beta +1) (\beta +6)^2}
    \Bigg[\frac{4 \alpha  (\beta +6) \left((\beta +6) (4 \beta -1) \left(\dfrac{r}{\sqrt{\theta}}\right)^{\frac{16 \beta +6}{\beta +6}}-6 (\beta +1) (2 \beta -3) \left(\dfrac{r_0}{\sqrt{\theta}}\right)^{\frac{15 \beta }{\beta +6}+1}\right)}{8 \beta +3}\Bigg.\\
    \Bigg. &
    -\frac{9 (\beta +1)^2 (2 \beta -1) M_* \left(\dfrac{r_0}{\sqrt{\theta}}\right)^{\frac{18 (\beta +1)}{\beta +6}} E_{\frac{45}{\beta +6}-8}\left(\frac{r_0^2/\theta}{4}\right)}{\pi ^{3/2}}\Bigg] +\frac{3 (2 \beta -1) M_* \dfrac{r^2}{\theta} E_{-\frac{9 (\beta +1)}{\beta +6}}\left(\frac{r^2/\theta}{4}\right)}{16 \pi ^{3/2} (\beta +6)},
    \end{split}
\end{equation}

\begin{equation}
    \begin{split}
      \Delta_L=&\frac{\frac{\alpha  \left(18 (\beta +1) (2 \beta -3) \left(\dfrac{r}{\sqrt{\theta}}\right)^{\frac{90}{\beta +6}} \left(\dfrac{r_0}{\sqrt{\theta}}\right)^{\frac{15 \beta }{\beta +6}+1}+(\beta +6) (4 \beta +9) \frac{r^{16}}{\theta^8}\right)}{(\beta +1) (8 \beta +3) \frac{r^{18}}{\theta^9}}-\frac{2 (7 \beta +3) M_*}{\pi ^2 \left(\frac{r^2}{\theta}+1\right)^2}}{\beta +6} \\
      &+\frac{6 (2 \beta -1) M_* \Gamma \left(\frac{5 (2 \beta +3)}{\beta +6}\right)} 
      {\pi ^2 (8 \beta +3) \frac{r^{18}}{\theta^9}}
      \Bigg[\frac{r^{16}}{\theta^8} \left(\, _2\tilde{F}_1\left(1,\frac{8 \beta +3}{\beta +6};\frac{9 (\beta +1)}{\beta +6};-\frac{r^2}{\theta}\right)-\, _2\tilde{F}_1\left(2,\frac{8 \beta +3}{\beta +6};\frac{9 (\beta +1)}{\beta +6};-\frac{r^2}{\theta}\right)\right)\Bigg.\\
      \Bigg.&
      +\left(\dfrac{r}{\sqrt{\theta}}\right)^{\frac{90}{\beta +6}} \left(\dfrac{r_0}{\sqrt{\theta}}\right)^{\frac{15 \beta }{\beta +6}+1} \Bigg(\, _2\tilde{F}_1\left(2,\frac{8 \beta +3}{\beta +6};\frac{9 (\beta +1)}{\beta +6};-\frac{r_0^2}{\theta}\right)
      -\, _2\tilde{F}_1\left(1,\frac{8 \beta +3}{\beta +6};\frac{9 (\beta +1)}{\beta +6};-\frac{r_0^2}{\theta}\right)\Bigg)\Bigg].
    \end{split}
\end{equation}
Figure \ref{fig:anisotropy} depicts the effect of anisotropy for a viable wormhole model under Gaussian and Lorentzian distributions. The investigation reveals that our anisotropy factor $\Delta$ is positive (i.e., $p_{t_*}>p_{r_*}$) and the structure of the wormhole is repulsive in Gaussian distribution [Figure \ref{fig:gaa}], whereas $\Delta$ is negative  (i.e., $p_{t_*}<p_{r_*}$) which indicates an attractive geometry of the wormhole in Lorentzian distribution [Figure \ref{fig:laa}].

%%%%%%%%%%%%%%%%%%%%%%%%%%%%%%%%%%%%%%%%%%%%%%%%%%%%%%%%%%%%%%%%%%%%%%%%%%%
\section{Results and Concluding Remarks}\label{section VI}
In this manuscript, we have explored the conformal symmetric wormhole solutions under non-commutative geometry in the background of $\mathpzc{f}(\mathcal{Q}, \mathcal{T})$ gravity. To achieve this, we have considered the presence of an anisotropic fluid in a spherically symmetric space-time. The concept of conformal symmetry and non-commutative geometry have already been used in literature within various context of modified theories of gravity \cite{aschieri,schne,sushkov,high,garatt,class,energy,cosmo,math,rahaman2,rahaman3,zinnat,kavya, kavya2}. Non-commutative geometry is used to replace the particle-like structure to smeared objects in string theory. Furthermore, conformal killing vectors are derived from the killing equation, which is based on the Lie algebra. These vectors are used to reduce the nonlinearity order of the field equation. Conformal symmetry has proved to be effective in describing relativistic stellar-type objects. Furthermore, it has led to new solutions and provided insights into geometry and kinematics \cite{pkfk}. It influences the geometry and dynamics of the spacetime, impacting key parameters such as throat size and stability.

In the framework of extended symmetric teleparallel gravity, we have derived some new exact solutions for wormholes by using both Gaussian and Lorentzian energy densities of non-commutative geometry. For this object, we presumed the linear wormhole model as $\mathpzc{f}(\mathcal{Q}, \mathcal{T}) =\alpha \mathcal{Q} + \beta \mathcal{T}$, where $\alpha$ and $\beta$ are model parameters. In both cases, we examined the wormhole scenario using Gaussian and Lorentzian distributions. By applying the throat condition in two distributions, we obtained different shape functions that obey all the criteria for a traversable wormhole. A similar result was presented in \cite{kavya} where the authors explored wormhole solutions in curvature matter coupling gravity supported by non-commutative geometry and conformal symmetry. Furthermore, we investigated the impact of model parameters on these two shape functions. Due to the conformal symmetry, the redshift function does not approach to zero as $r >r_0$ \cite{kuhfitting, rahaman2, rahaman3, shahz}.

Figures \ref{fig:gsf} and \ref{fig:lsf} show the graphical nature of the obtained shape functions with $\beta \in [0, 0.5)$. Notably, a slight variation in the value of $\beta$ can impact the nature of the shape function. Moreover, the graphical behavior of energy conditions are shown in Figures \ref{fig:gec} and \ref{fig:lec}. The energy density is positive throughout the space-time. For all the wormhole solutions, the violation of the null energy conditions indicates the presence of hypothetical matter. Here, this nature of hypothetical fluid presented in references \cite{kavya,zn1,mg1}. Next, we studied the effect of anisotropy for both distributions. The geometry of the wormhole is repulsive in Gaussian distribution, whereas attractive in Lorentzian distribution [Figure \ref{fig:anisotropy}]. 

To conclude, this work validates the conformal symmetric wormhole solutions in $\mathpzc{f}(\mathcal{Q}, \mathcal{T})$ gravity under non-commutative geometry. The authors \cite{gm4} have identified the possibility of generalized wormhole formation in the galactic halo due to dark matter using observational data within the matter coupling gravity formalism. In near future, we plan to investigate various wormhole scenarios in alternative theories of gravity, as discussed in references \cite{gm5,gm6,gm7,gm8}.
%%%%%%%%%%%%%%%%%%%%%%%%%%%%%%%%%%%%%%%%%%%%%%%%%%%%%%%%%%%%%%%%%%%%%%%%%%%%
\section*{Data Availability Statement}
There are no new data associated with this article.

\begin{acknowledgments}
  C.C.C.,  V.V. and N.S.K. acknowledge DST, New Delhi, India, for its financial support for research facilities under DST-FIST-2019. 
\end{acknowledgments}

%\nocite{*}

\end{document}